\documentclass[preprint,showpacs,preprintnumbers,amsmath,amssymb]{revtex4}
\usepackage{booktabs}
\usepackage{mathrsfs}
\usepackage{epsfig}
\usepackage{graphicx}
\usepackage{dcolumn}
\usepackage{bm}
\usepackage{amsmath}
\usepackage{slashed}       

\let\jnfont=\rm
\def\NPB#1,{{\jnfont Nucl.\ Phys.\ B }{\bf #1},}
\def\PLB#1,{{\jnfont Phys.\ Lett.\ B }{\bf #1},}
\def\EPJC#1,{{\jnfont Eur.\ Phys.\ Jour.\ C }{\bf #1},}
\def\PRD#1,{{\jnfont Phys.\ Rev.\ D }{\bf #1},}
\def\PRL#1,{{\jnfont Phys.\ Rev.\ Lett.\ }{\bf #1},}
\def\MPLA#1,{{\jnfont Mod.\ Phys.\ Lett.\ A }{\bf #1},}
\def\JPG#1,{{\jnfont J.\ Phys.\ G}{\bf #1},}
\def\CTP#1,{{\jnfont Commun.\ Theor.\ Phys.\ }{\bf #1},}
\def\ZPC#1,{{\jnfont Z.\ Phys.\ C }{\bf #1},}
\def\JHEP#1,{{\jnfont JHEP \ }{\bf #1},}
\def\lsim{\raise0.3ex\hbox{$<$\kern-0.75em\raise-1.1ex\hbox{$\sim$}}}
\def\gsim{\raise0.3ex\hbox{$>$\kern-0.75em\raise-1.1ex\hbox{$\sim$}}}

\begin{document}

\title{A 125 GeV Higgs and its di-photon signal in different SUSY models: a mini review}

\author{Zhaoxia Heng}

\affiliation{ Department of Physics,
       Henan Normal University, Xinxiang 453007, China
      \vspace{1cm}}

\begin{abstract}
In this note we briefly review our recent studies on a 125 GeV Higgs and its di-photon signal rate
in different low energy supersymmetric models, namely the minimal supersymmetric
standard model (MSSM), the next-to-minimal supersymmetric standard model (NMSSM),
the nearly minimal supersymmetric standard model (nMSSM) and the constrained MSSM.
Our conclusion is: (i) In the allowed parameter space the SM-like Higgs boson can easily
be 125 GeV in the MSSM, NMSSM and nMSSM, while it is hard to realize in the constrained MSSM;
(ii) The di-photon Higgs signal rate in the nMSSM and constrained MSSM is
suppressed relative to the prediction of the SM, while the signal rate
can be enhanced in the MSSM and NMSSM; (iii) The NMSSM may allow for a lighter
top-squark than the MSSM, which can thus ameliorate the fine-tuning problem.
\end{abstract}

\maketitle

\section{Introduction}
Considering the important role of the Higgs boson in particle physics,
hunting for it has been one of the major tasks of the running Large Hadron Collider (LHC).
Recently, both the ATLAS and CMS collaborations have reported some evidence for a light Higgs boson
near 125 GeV \cite{ATLAS,CMS} with a di-photon signal rate slightly above the SM prediction \cite{Carmi}.

As is well known, in new physics beyond the SM model several Higgs bosons are predicted,
among which the SM-like one may be near 125 GeV \cite{125Higgs,125other,cao125,Cao:2011sn}.
Recently, in our studies \cite{cao125,Cao:2011sn}
we examined the mass of the SM-like Higgs boson in several supersymmetric (SUSY) models including
the minimal supersymmetric standard model (MSSM)\cite{Haber,Djouadi}, the next-to-minimal supersymmetric
standard model (NMSSM)\cite{NMSSM1,NMSSM2} and the constrained MSSM\cite{mSUGRA,nuhm2}.
At tree-level, these SUSY models are hard to predict a Higgs boson near 125 GeV,
and sizable radiative corrections, which mainly come from the top and top-squark loops,
are necessary to enhance the Higgs boson mass\cite{Carena:1995bx}. Due to the different
properties of these SUSY models, the loop contributions to the Higgs boson mass are different
for giving a 125 GeV Higgs boson. Therefore, different models have different lower bounds on
the top-squark mass which is associated with the fine-tuning problem \cite{tuning}.
On the other hand, since the di-photon Higgs signal is the most promising discovery
channel for a light Higgs boson at the LHC\cite{diphoton1}, in our recent study \cite{di-photon} we
performed a comparative study for the di-photon Higgs signal in different SUSY models,
namely the MSSM, NMSSM and the nearly mininal supersymmertric
standard model (nMSSM) \cite{xnMSSM,cao-xnmssm}.
In this note we briefly review these stuides on
a 125 GeV Higgs boson and its di-photon signal rate
in different SUSY models.

This note is organized as follows. In the next section we briefly describe the Higgs sector
and the di-photon Higgs signal in these SUSY models. Then we present the numerical results
and discussions in Sec. III. Finally, the conclusions are given in Sec. IV.

\section{The Higgs sector and di-photon signal rate in SUSY models}
\subsection{The Higgs sector in SUSY models}
Different from the SM, the Higgs sector in the supersymmetric models
is usually extended by adding Higgs doublets and/or singlets.
The most economical realization is the MSSM, which consists of
two Higgs doublet $H_u$ and $H_d$. In order to solve the $\mu-$problem
and the little hierarchy problem in the MSSM, the singlet extension of MSSM,
such as the NMSSM\cite{NMSSM1} and nMSSM\cite{xnMSSM,cao-xnmssm}
has been intensively studied\cite{Barger}.
The differences between these models come from their
superpotentials and the corresponding soft-breaking terms, which are given by:
\begin{eqnarray}
 W_{\rm MSSM}&=& W_F + \mu \hat{H_u}\cdot \hat{H_d}, \label{MSSM-pot}\\
 W_{\rm NMSSM}&=&W_F + \lambda\hat{H_u} \cdot \hat{H_d} \hat{S}
 + \frac{1}{3}\kappa \hat{S^3},\\
  W_{\rm nMSSM}&=&W_F + \lambda\hat{H_u} \cdot \hat{H_d} \hat{S}
  +\xi_F M_n^2\hat S,\\
 V_{\rm soft}^{\rm MSSM}&=&\tilde m_u^2|H_u|^2 + \tilde m_d^2|H_d|^2
+ (B\mu H_u\cdot H_d + h.c.),\\
V_{\rm soft}^{\rm NMSSM}&=&\tilde m_u^2|H_u|^2 + \tilde m_d^2|H_d|^2
+ \tilde m_S^2|S|^2 +(A_\lambda \lambda SH_u\cdot H_d
+\frac{A_\kappa}{3}\kappa S^3 + h.c.),\\
V_{\rm soft}^{\rm nMSSM}&=&\tilde m_u^2|H_u|^2 + \tilde m_d^2|H_d|^2
+ \tilde m_S^2|S|^2 +(A_\lambda \lambda SH_u\cdot H_d
+\xi_S M_n^3 S + h.c.),
\end{eqnarray}
where $W_F$ is the MSSM superpotential without the $\mu$ term,
$\lambda$, $\kappa$ and $\xi_F$ are the dimensionless parameters and
$\tilde{m}_{u}$, $\tilde{m}_{d}$, $\tilde{m}_{S}$, $B$,
$A_\lambda$, $A_\kappa$ and $\xi_S M_n^3$ are soft-breaking parameters.
Note that in the NMSSM and nMSSM the $\mu$-term is replaced by the
$\mu_{\rm eff}=\lambda s$ when the singlet Higgs field $\hat S$ develops
a VEV $s$. The differences between the NMSSM and nMSSM reflect the last term
in the superpotential, where the cubic singlet term $\kappa \hat{S}^3$ in the NMSSM
is replaced by a tadpole term $\xi_F M_n^2 \hat{S}$ in the nMSSM.
This replacement in the superpotential makes the nMSSM has no discrete symmetry and
thus free of the domain wall problem that the NMSSM suffers from.
Actually, due to the tadpole term $\xi_F M_n^2$ does not
induce any interaction, the nMSSM is identical to the NMSSM with $\kappa=0$,
except for the minimization conditions of the Higgs potential and the tree-level
Higgs mass matrices.

With the superpotentials and the soft-breaking terms giving above, one
can get the Higgs potentials of these SUSY models, and then
can derive the Higgs mass matrics and eigenstates.
At the minimum of the potential, the Higgs fields $H_u$, $H_d$ and $S$ are expanded as
\begin{eqnarray}
H_u = \left ( \begin{array}{c} H_u^+ \\
       v_u +\frac{ \phi_u + i \varphi_u}{\sqrt{2}}
        \end{array} \right),~~
H_d & =& \left ( \begin{array}{c}
             v_d + \frac{\phi_d + i \varphi_d}{\sqrt{2}}\\
             H_d^- \end{array} \right),~~
S  =  s + \frac{1}{\sqrt{2}} \left(\sigma + i \xi \right),
\end{eqnarray}
with $v=\sqrt{v_u^2+v_d^2}=$ 174 GeV.
By unitary rotation the mass eigenstates can be given by
\begin{eqnarray} \left( \begin{array}{c} h_1 \\
h_2 \\ h_3 \end{array} \right) = S \left( \begin{array}{c} \phi_u
\\ \phi_d\\ \sigma\end{array} \right),~ \left(\begin{array}{c} a\\
A\\ G^0 \end{array} \right) = P \left(\begin{array}{c} \varphi_u
\\ \varphi_d \\ \xi \end{array} \right),~ \left(\begin{array}{c} H^+
\\G^+ \end{array}  \right) =U \left(\begin{array}{c}H_u^+\\ H_d^+
\end{array} \right).  \label{rotation}
\end{eqnarray}
where $h_1,h_2,h_3$ are physical CP-even Higgs bosons ($m_{h_1}<m_{h_2}<m_{h_3}$), $a,A$ are CP-odd
 Higgs bosons, $H^+$ is the charged Higgs boson, and $G^0$, $G^+$
are Goldstone bosons eaten by $Z$ and $W^+$. Due to the absence of
the singlet field $S$, the MSSM only has two CP-even Higgs bosons
and one CP-odd Higgs bosons, as well as one pair of charged Higgs bosons.

At the tree-level, the Higgs masses in the MSSM are conventionally
parameterized in terms of the mass of the CP-odd Higgs boson ($m_A$) and $\tan\beta\equiv v_u/v_d$
and the loop corrections typically come from top and
stop loops due to their large Yukawa coupling. For small splitting between the
stop masses, an approximate formula of the lightest Higgs boson mass is given by\cite{Carena:2011aa},
\begin{equation}\label{mh}
 m^2_{h}  \simeq M^2_Z\cos^2 2\beta +
  \frac{3m^4_t}{4\pi^2v^2} \ln\frac{M^2_S}{m^2_t} +
\frac{3m^4_t}{4\pi^2v^2}\frac{X^2_t}{M_S^2} \left( 1 - \frac{X^2_t}{12M^2_S}\right),
\end{equation}
where $M_S = \sqrt{m_{\tilde{t}_1}m_{\tilde{t}_2}}$ and
$X_t \equiv A_t - \mu \cot \beta$.
The formula manifests that larger $M_S$ or $\tan\beta$
is necessary to push up the Higgs boson mass. And the Higgs boson mass can reach a maximum
when $X_t/M_S=\sqrt{6}$ for given $M_S$ (i.e. the so-called $m_h^{max}$ scenario).
Note that the lightest Higgs boson is the SM-like Higgs boson $h$ (with the largest
coupling to vector bosons) in most of the MSSM parameter space.

Different from the case in the MSSM, the Higgs sector in the NMSSM depends on the following
six parameters,
\begin{eqnarray}
\lambda, \quad \kappa, \quad M_A^2= \frac{2 \mu (A_\lambda + \kappa s)}{\sin 2 \beta},
\quad A_\kappa, \quad \tan \beta=\frac{v_u}{v_d}, \quad \mu = \lambda s.
\end{eqnarray}
and in the nMSSM the input parameters in the Higgs sector are
\begin{eqnarray}
\lambda, \quad \tan\beta, \quad \mu \quad A_\lambda, \quad \tilde m_S,
 \quad M_A^2=\frac{2(\mu A_\lambda + \lambda \xi_F M_n^2)}{\sin 2 \beta}.
\end{eqnarray}
Because the coupling $\lambda\hat{H_u} \cdot \hat{H_d} \hat{S}$ in the superpotential,
the tree-level Higgs boson mass has an additional contribution in the NMSSM and nMSSM,
\begin{eqnarray}
\Delta m_h^2= \lambda^2 v^2 \sin^2 2\beta
\end{eqnarray}
In order to push up the tree-level Higgs boson mass,
$\lambda$ has to be as large as possible and $\tan\beta$
has to be small. The requirement of the absence of a landau singularity
below the GUT scale implies that $\lambda\lesssim$ 0.7 at the weak scale,
and the upper bound on $\lambda$ at the weak scale depends strongly on
$\tan\beta$ and grows with increasing $\tan\beta$\cite{king}.
However, this can still lead to
a larger tree-level Higgs boson mass than in the MSSM.
Therefore, the radiative corrections to $m_h^2$ may be reduced in the NMSSM and nMSSM,
which may induce light top-squark and ameliorate the fine-tuning problem\cite{tuning2}.

In the NMSSM and nMSSM, due to the mixing between the doublet Higgs fields and the
singlet Higgs field, the SM-like Higgs boson $h$ may either be the lightest CP-even
Higgs boson or the next-to-lightest CP-even Higgs boson, which corresponds to the so-called
pull-down case or the push-up case\cite{cao125}, respectively.
Although the mass of the SM-like Higgs boson in the nMSSM is quite similar to that
in the NMSSM, the Higgs signal is quite different.
This is because the peculiarity of the neutralino sector in the nMSSM, where
the lightest neutralino $\tilde{\chi}^0_1$ as the lightest supersymmetric particle(LSP)
acts as the dark matter candidate, and its mass takes the form\cite{rarez}
\begin{eqnarray}
m_{\tilde{\chi}^0_1} \simeq \frac{2\mu \lambda^2 v^2}{\mu^2+\lambda^2 v^2}
             \frac{\tan \beta}{\tan^2 \beta+1}
\end{eqnarray}
This expression implies that $\tilde{\chi}_1^0$ must be lighter than
about $60$ GeV for $\lambda < 0.7$ (perturbativity bound) and
$\mu > 100 {\rm GeV}$ (from lower bound on chargnio mass).
And $\tilde{\chi}^0_1$ must annihilate by exchanging a resonant light
CP-odd Higgs boson to get the correct relic density.
For such a light neutralino, the SM-like Higgs boson around 125GeV tends
to decay predominantly into light neutralinos or other light Higgs bosons\cite{cao-xnmssm}.

\subsection{The di-photon Higgs signal}
Considering the di-photon signal is of prime importance to searching for Higgs boson near 125 GeV,
it is necessary to estimate its signal rate, and we define the normalized
production rate as
\begin{eqnarray}
R_{\gamma\gamma} &\equiv & \sigma_{SUSY} ( p p \to h \to
\gamma \gamma)/\sigma_{SM} ( p p \to h \to \gamma \gamma ) \nonumber \\
&\simeq& [\Gamma(h\to gg) Br(h\to \gamma\gamma)] /[\Gamma(h_{SM}\to
gg) Br(h_{SM}\to \gamma\gamma)] \nonumber \\
&=&  [\Gamma(h\to gg) \Gamma(h\to \gamma\gamma)] /[\Gamma(h_{SM}\to
gg) \Gamma(h_{SM}\to \gamma\gamma)] \times
\Gamma_{tot}(h_{SM})/\Gamma_{tot}(h) \nonumber\\
&=& C_{hgg}^2 C_{h\gamma\gamma}^2\times
\Gamma_{tot}(h_{SM})/\Gamma_{tot}(h) \label{definition}
\end{eqnarray}
where $C_{hgg}$ and $C_{h\gamma\gamma}$ are the couplings of Higgs
 to gluons and photons in SUSY with respect to their SM values, respectively.
In SUSY, the $hgg$ coupling arises mainly from the loops mediated by the third
generation quarks and squarks, while the $h\gamma\gamma$ coupling has additional
contributions from loops mediated by W-boson, charged Higgs boson, charginos and
the third generation leptons and sleptons. Their decay widths are given by\cite{Djouadi}
\begin{eqnarray}
 \Gamma(h\to gg)&=&\frac{G_F \alpha_s^2 m_h^3}{36 \sqrt{2}\pi^3}
       \left| \frac{3}{4}\sum_q g_{hqq}\, A_{1/2}^h(\tau_q)
       +\frac{3}{4}{\cal A}^{gg}
     \right|^2 \label{hgg}\\
 \Gamma(h\to \gamma\gamma)&=&\frac{G_{F}\alpha^{2}m_{h}^{3}}{128\sqrt{2}\pi}
\left| \sum_f N_{c}\, Q_{f}^{2}\, g_{hff}\, A_{1/2}^{h}(\tau_{f})+
g_{hWW}\, A_{1}^{h}(\tau_{W}) + {\cal A}^{\gamma\gamma}\right|^2, \label{hgaga}
\end{eqnarray}
with $\tau_i = m_h^2/(4m_i^2)$, and
\begin{eqnarray}
{\cal A}^{gg} &=&  \sum_{i}
 \frac{g_{h\tilde{q}_i\tilde{q}_i}}{m_{\tilde{q}_i}^2}A_{0}^h(\tau_{\tilde{q}_i}),\nonumber\\
{\cal A}^{\gamma\gamma} &=&
 g_{hH^{+}H^{-}}\frac{m_{W}^{2}}{m^{2}_{H^{\pm}}}A_{0}^{h}(\tau_{H^{\pm}})
 +\sum_{f}\frac{g_{h\tilde{f}\tilde{f}}}{m_{\tilde{f}}^2}A_{0}^h(\tau_{\tilde{f}})
 +\sum_i g_{h\chi_i^+\chi_i^-}\frac{m_W}{m_{\chi_i}} A_{1/2}^h(\tau_{\chi_i}) \label{agg},
\end{eqnarray}
where $m_{\tilde{f}}$ and $m_{\chi_i}$ are the mass of sfermion and chargino, respectively.
In the limit $\tau_i \ll 1$, the asymptotic behaviors of $A_i^h$ are given by
\begin{equation}\label{asymp}
A_0^h \to - \frac13\ , \quad A_{1/2}^h \to -\frac43\ ,\quad
A_{1}^h \to +7 \ ,
\end{equation}
One can easily learn that the W-boson contribution to $h\gamma\gamma$ is
by far dominant, however, for light stau or squarks with large mixing,
the $h\gamma\gamma$ coupling can be enhanced. While light squarks with
large mixing can suppress the $hgg$ coupling. Therefore, light stau with
large mixing may enhance the di-photon signal rate\cite{Carena:2011aa}, while light squarks
with large mixing have little effect on the di-photon signal rate.

\section{Numerical Results and discussions}

In our work the Higgs boson mass are calculated by the package
 NMSSMTools\cite{NMSSMTools}, which include the dominant one-loop
and leading logarithmic two-loop corrections.
Considering the Higgs hints at the LHC, we focus on the Higgs boson
mass between 123 GeV and 127 GeV, furthermore, we consider the
following constraints:
\begin{itemize}
\item [(1)]The constraints from LHC experiment for the non-standard Higgs boson.
\item [(2)]The constraints from LEP and Tevatron on the masses
of the Higgs boson and sparticles, as well as on the neutralino pair productions.
\item [(3)]The indirect constraints from B-physics (such as the latest experimental result
of $B_s\to \mu^+\mu^-$) and from the electroweak
precision observables such as $M_W$, $\sin^2 \theta_{eff}^{\ell}$ and $\rho_{\ell}$,
or their combinations $\epsilon_i (i=1,2,3)$ \cite{Altarelli}.
\item [(4)]The constraints from the muon $g-2$:
$a_\mu^{exp}-a_\mu^{SM} = (25.5\pm 8.2)\times 10^{-10}$ \cite{g-2}. We require
SUSY to explain the discrepancy at $2\sigma$ level.
\item [(5)]The dark matter constraints from WMAP relic density
(0.1053 $< \Omega h^2 <$ 0.1193) \cite{WMAP} and the direct detection exclusion limits
on the scattering cross section from XENON100 experiment (at $90\%$ C.L.) \cite{XENON}.
\end{itemize}
Note that most of the above constraints have been encoded in the package NMSSMTools.

\begin{figure}[t]
\centering
\includegraphics[width=7cm]{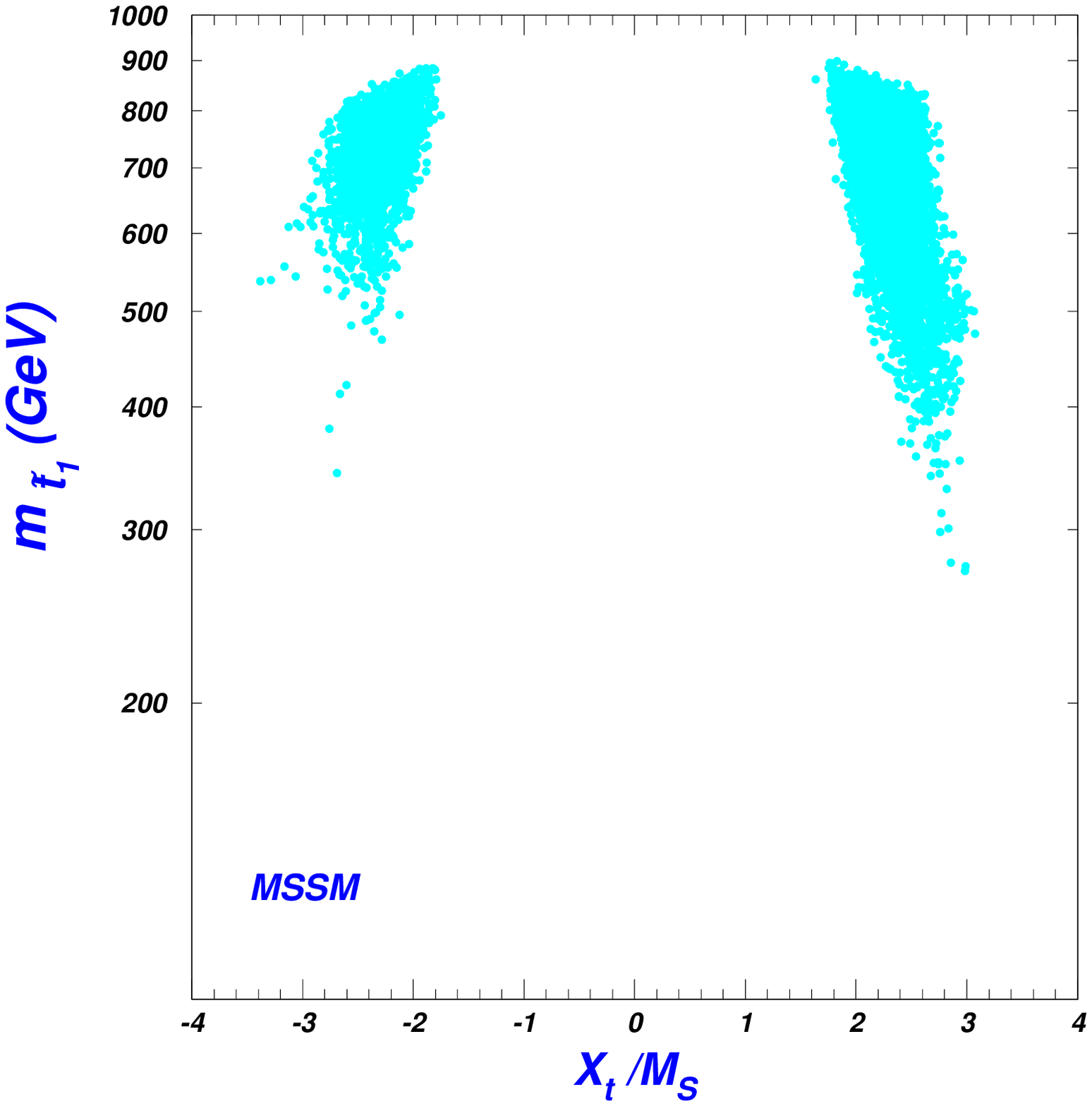}\hspace{0.2cm}
\includegraphics[width=7cm]{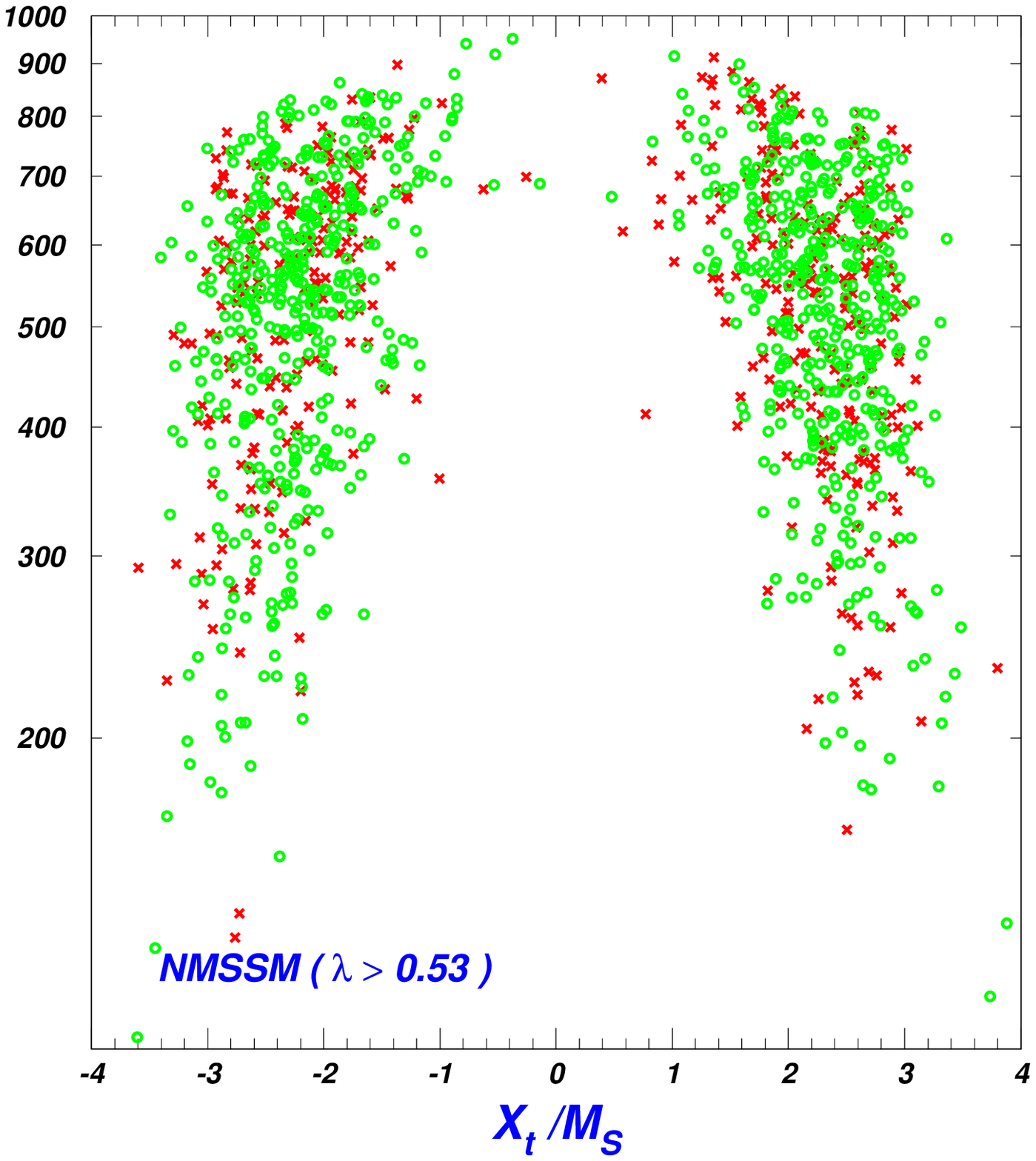}
\vspace*{-0.5cm}
\caption{The scatter plots of the samples in the MSSM and NMSSM (with
$\lambda>$ 0.53) satisfying various constraints listed in the text
(including $ 123{\rm GeV} \leq m_h \leq 127 {\rm GeV} $), showing
the correlation between the mass of the lighter top-squark and $X_t/M_{S}$
with $M_S \equiv \sqrt{m_{\tilde t_1} m_{\tilde t_2}}$ and $X_t \equiv A_t - \mu \cot \beta$.
In the right panel the circles (green) denote the pull-down case (the lightest Higgs boson
 being the SM-like Higgs), and the times (red) denote the push-up case
(the next-to-lightest Higgs boson being the SM-like Higgs).}
\label{fig1}
\end{figure}
Natural supersymmetry are usually characterized by a small superpotential parameter $\mu$,
and the third generation squarks with mass $\lesssim 0.5-1.5$ TeV\cite{naturalsusy}. Therefore,
we only consider the case with
\begin{eqnarray}
100{\rm ~GeV}\leq (M_{Q_3},M_{U_3})\leq 1 {\rm ~TeV} ,~~|A_{t}|\leq 3 {\rm ~TeV}.
\label{narrow}
\end{eqnarray}
For the case with $\lambda<0.2$ in the NMSSM, the property of the NMSSM is similar to
the case in the MSSM\cite{cao125}. In order to distinguish the features
between MSSM and NMSSM, we only consider the case with $\lambda> m_Z/v \simeq 0.53$ in
the NMSSM. We scan over the parameter space of the MSSM and NMSSM under
the above experimental constraints and study the property of the
Higgs boson for the samples surviving the constraints.

\begin{figure}[htbp]
\centering
\includegraphics[width=6.5cm]{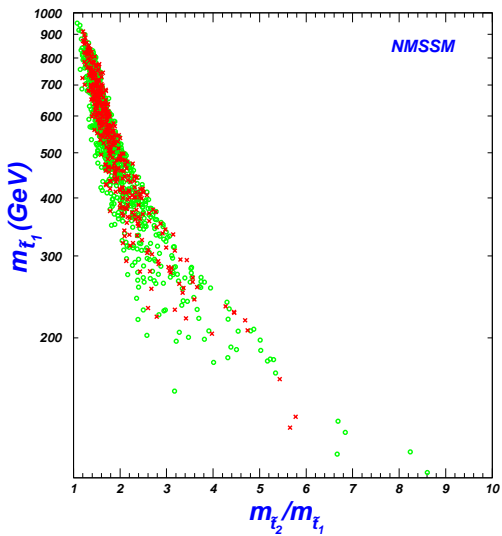}
\vspace*{-0.5cm}
\caption{Same as Fig.\ref{fig1}, but only for the NMSSM, showing the correlation between
 $m_{\tilde t_1}$ and the ratio $m_{\tilde t_2}/m_{\tilde t_1}$.}
\label{fig2}
\end{figure}

In Fig.\ref{fig1} we display the surviving samples in the MSSM and NMSSM (with $\lambda>$ 0.53),
showing the correlation between the lighter top-squark mass and the ratio $X_t/M_{S}$ with
$M_S \equiv \sqrt{m_{\tilde t_1} m_{\tilde t_2}}$. From the figure we see that
for a moderate light $\tilde t_1$, large $X_t$ is necessary to satisfy $m_h\sim$ 125 GeV,
and for large $m_{\tilde t_1}$, the ratio $X_t/M_{S}$ decreases.
In the MSSM, $|X_t/M_{S}|>$ 1.6 for $m_{\tilde t_1} <$ 1 TeV, i.e. no-mixing scenario($X_t=0$)
 cannot survive, and the top-squark mass is usually larger than 300 GeV.
This implies that a large top-squark mass or a near-maximal stop mixing
is necessary to satisfy the Higgs mass near 125 GeV.
However, the case is very different in the NMSSM, $X_t\approx 0$ may also survive,
and the lighter top-squark mass can be as light as about 100 GeV, which may alleviate the fine-tuning
problem and make the NMSSM seems more natural. In the case of light $m_{\tilde t_1}$,
$|X_t/M_{S}|$ is usually larger than $\sqrt{6}$, which corresponds to a large splitting between
$m_{\tilde t_1}$ and $m_{\tilde t_2}$, as the Fig.\ref{fig2} shown.

\begin{figure}[t]
\includegraphics[width=6.5cm]{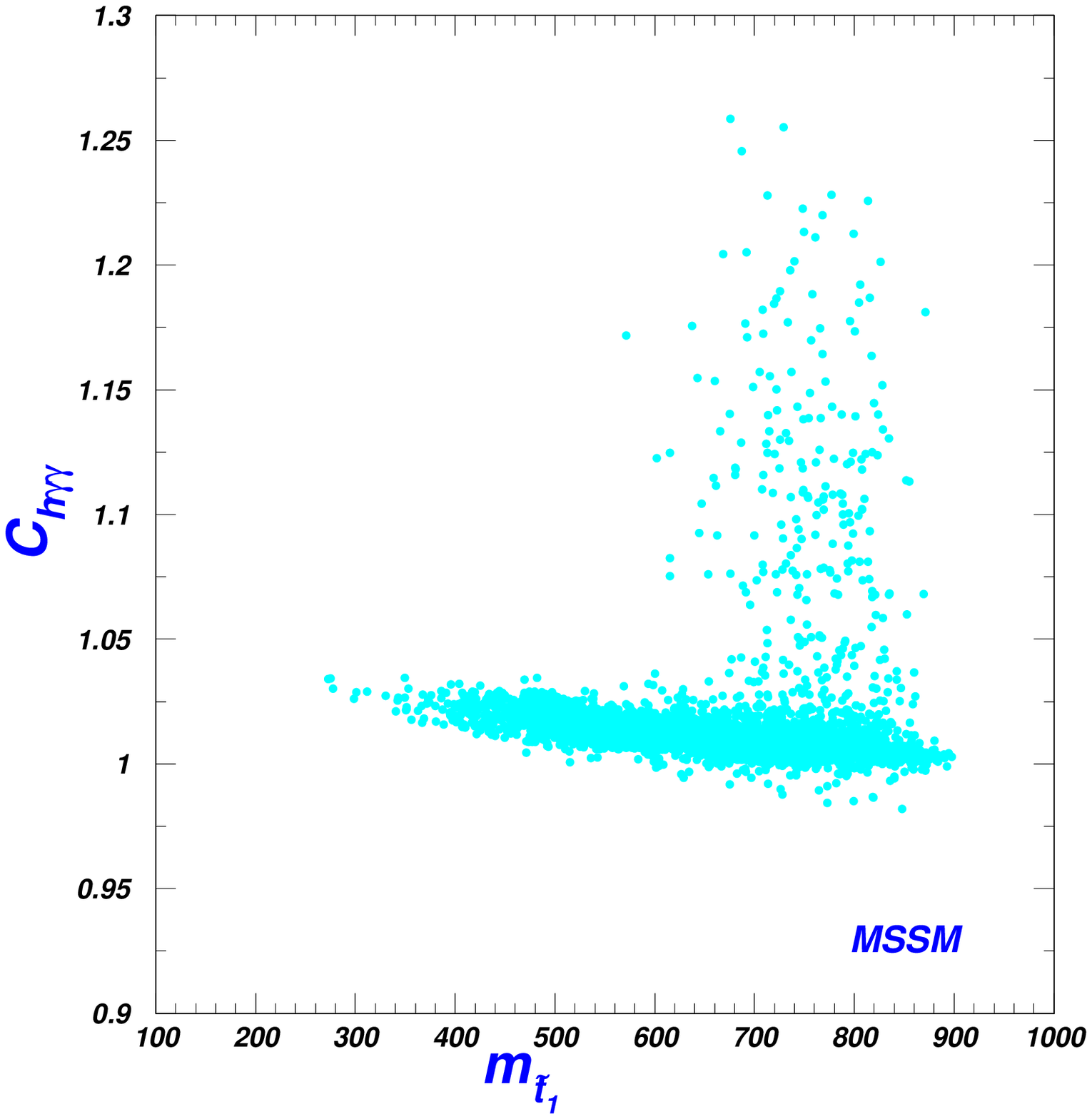}\hspace{0.2cm}
\includegraphics[width=6.5cm]{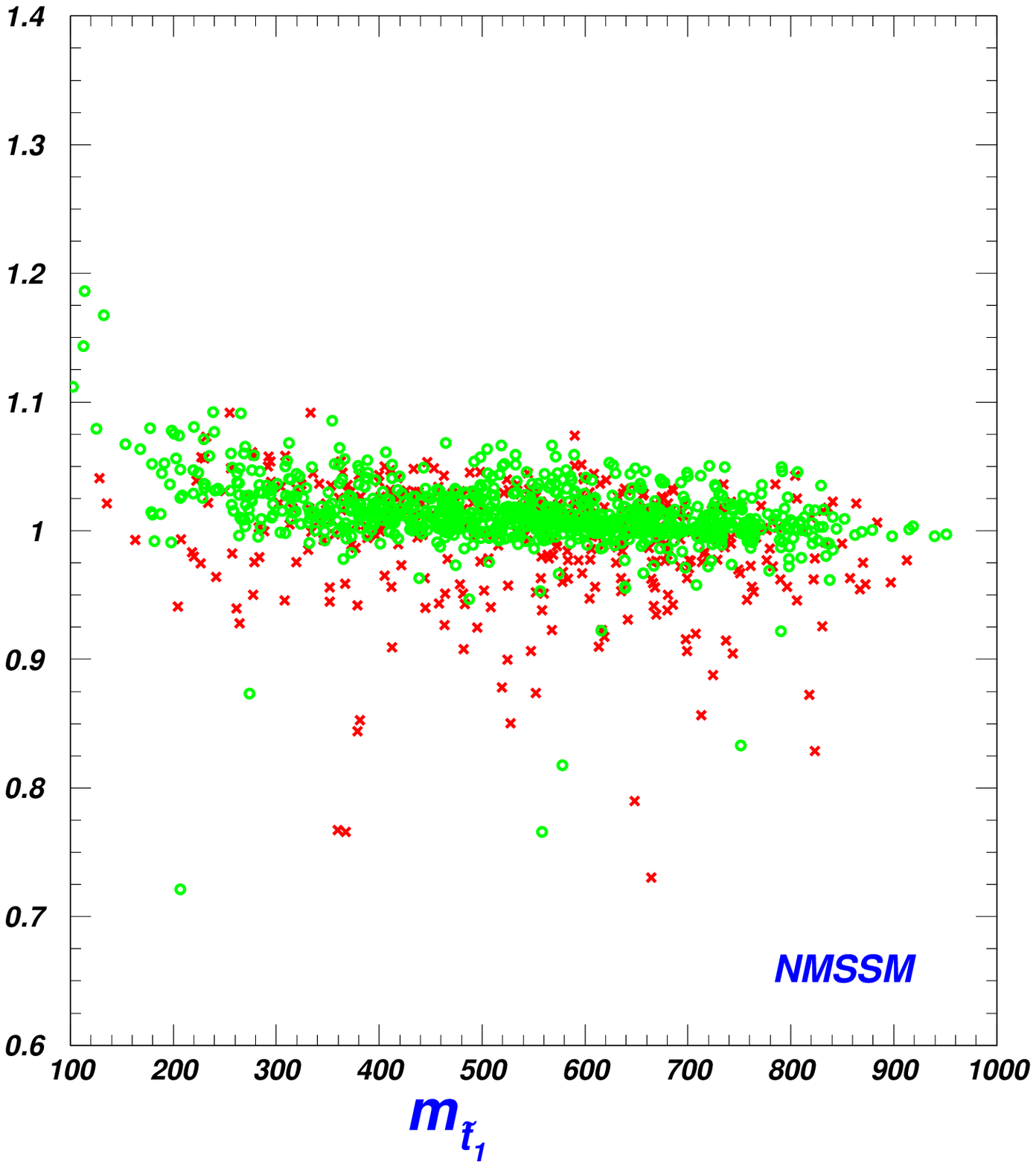}
    \includegraphics[width=6.5cm]{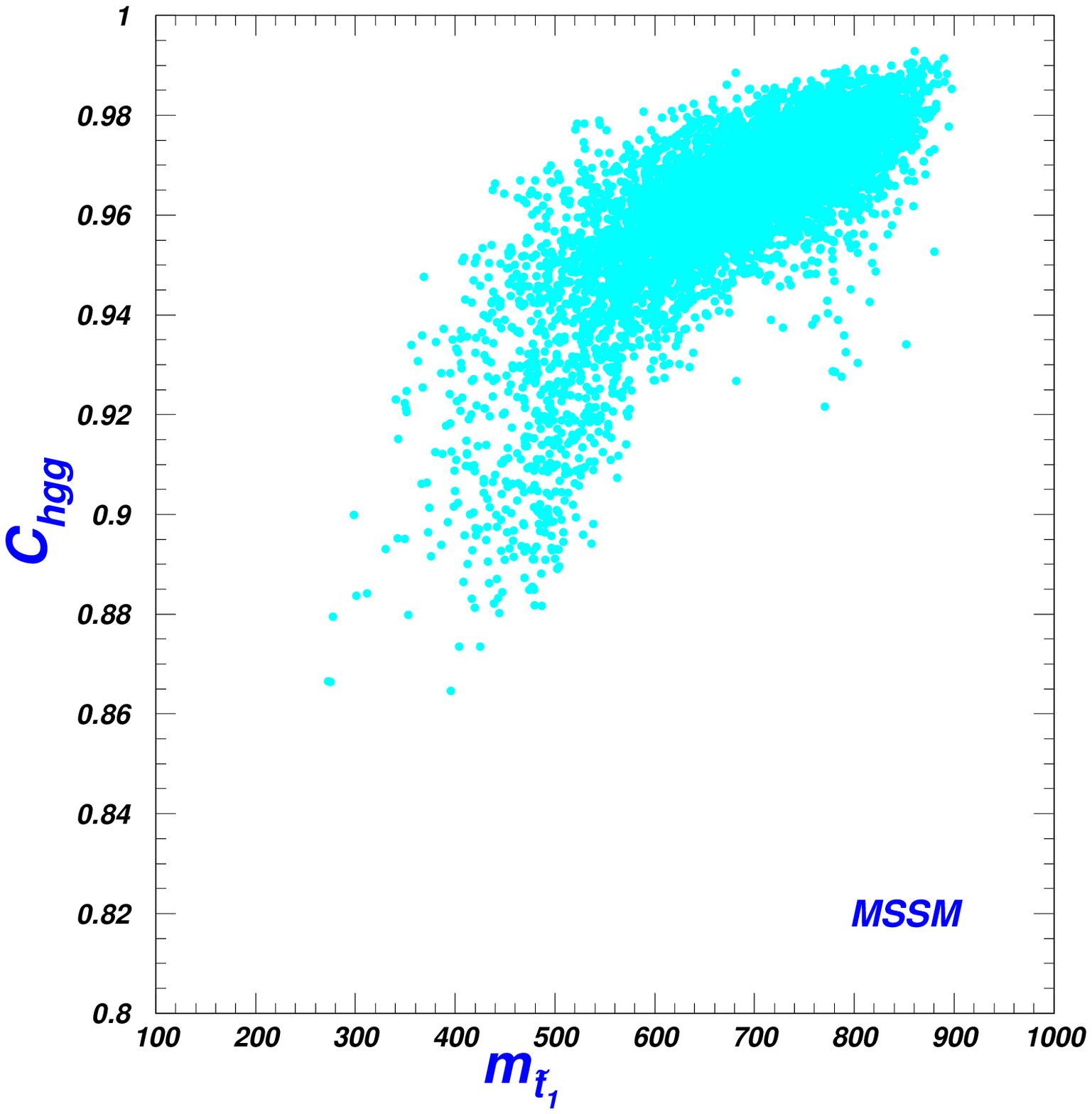}\hspace{0.2cm}
    \includegraphics[width=6.5cm]{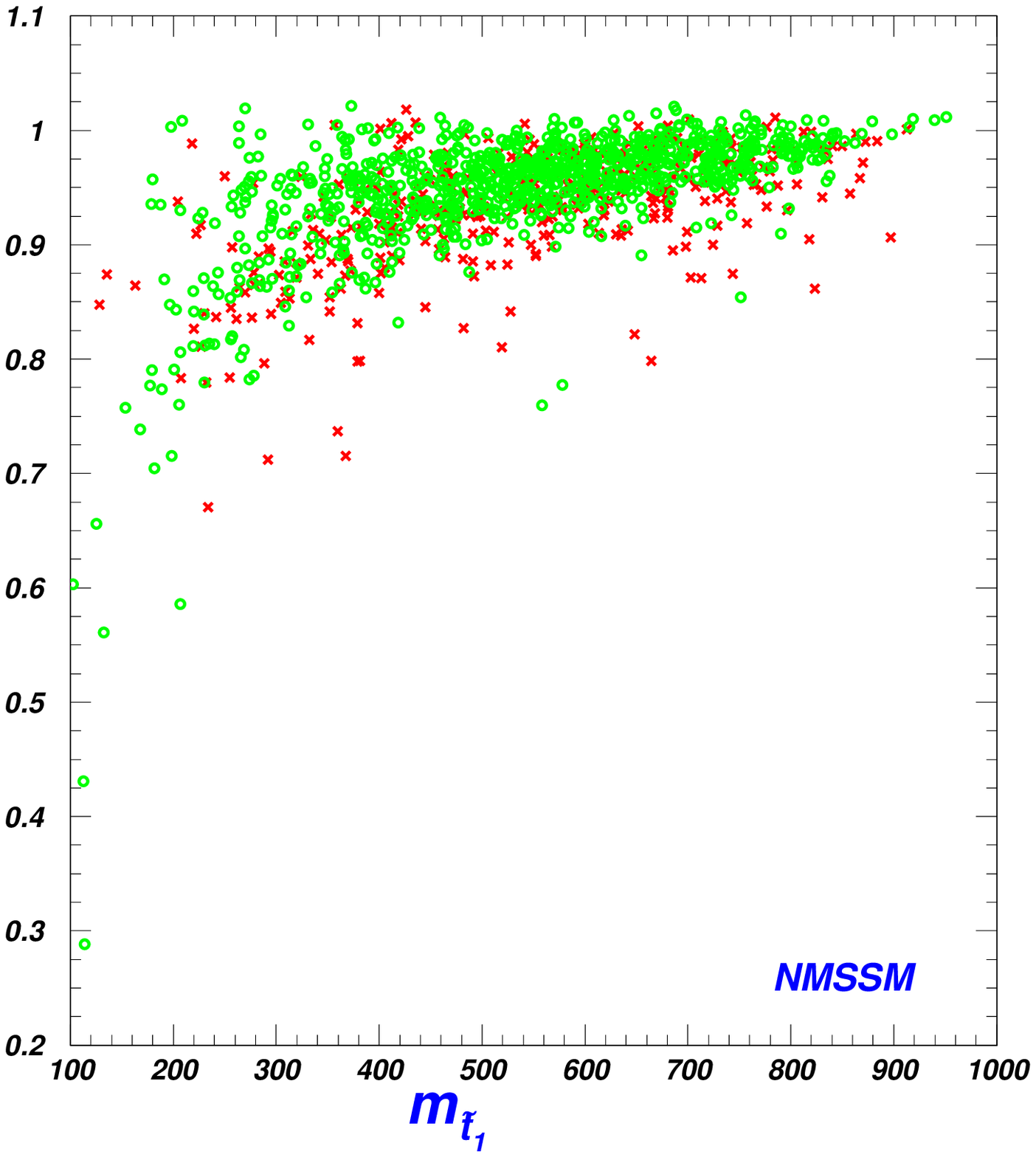}
\vspace*{-0.5cm}
\caption{Same as Fig.1, projected in the planes of $m_{\tilde t_1}$ versus the reduced
couplings $C_{h\gamma\gamma}$ and $C_{hgg}$, respectively. }
\label{fig3}
\end{figure}
\begin{figure}[htbp]
\centering
\includegraphics[width=7cm]{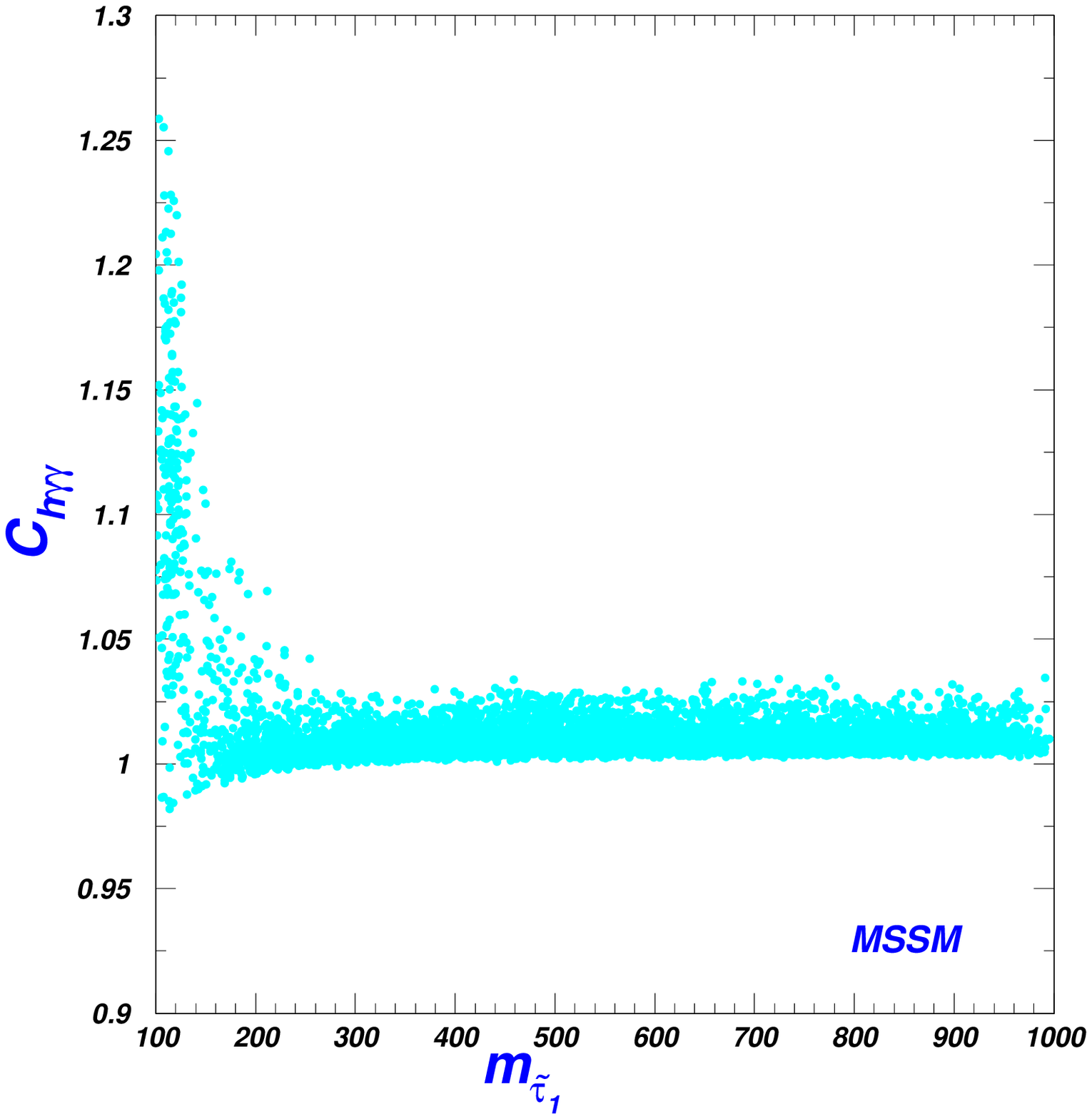}\hspace{0.2cm}
\includegraphics[width=7cm]{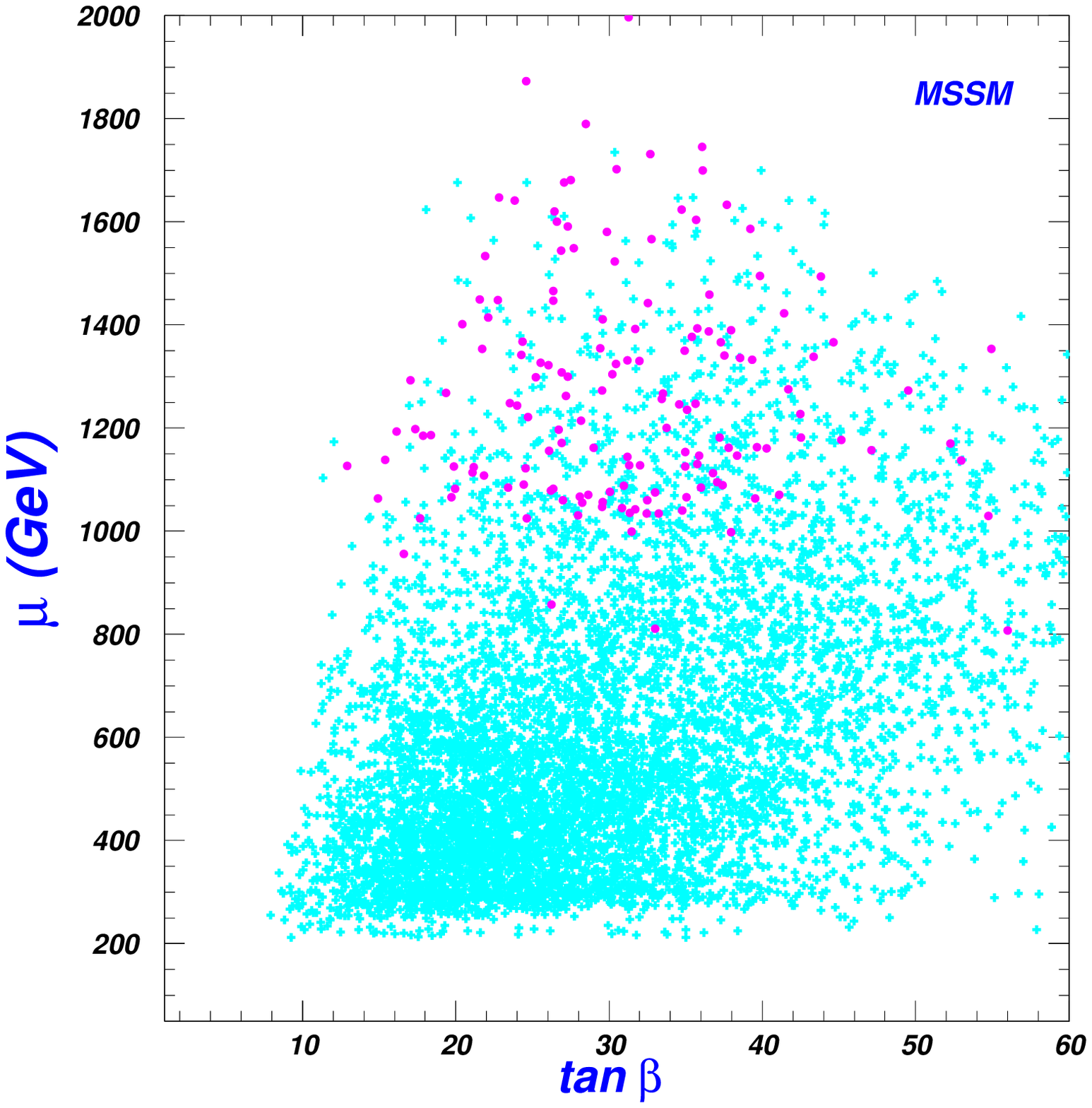}
\vspace*{-0.5cm}
\caption{Same as Fig.\ref{fig1}, but only for the MSSM, showing the correlation between
$m_{\tilde \tau_1}$ and the reduced coupling $C_{h\gamma\gamma}$, $\mu$ and $\tan\beta$,
respectively. The purple points correspond to $R_{\gamma\gamma}>1$.}
\label{fig4}
\end{figure}
\begin{figure}[htbp]
\centering
\includegraphics[width=7cm]{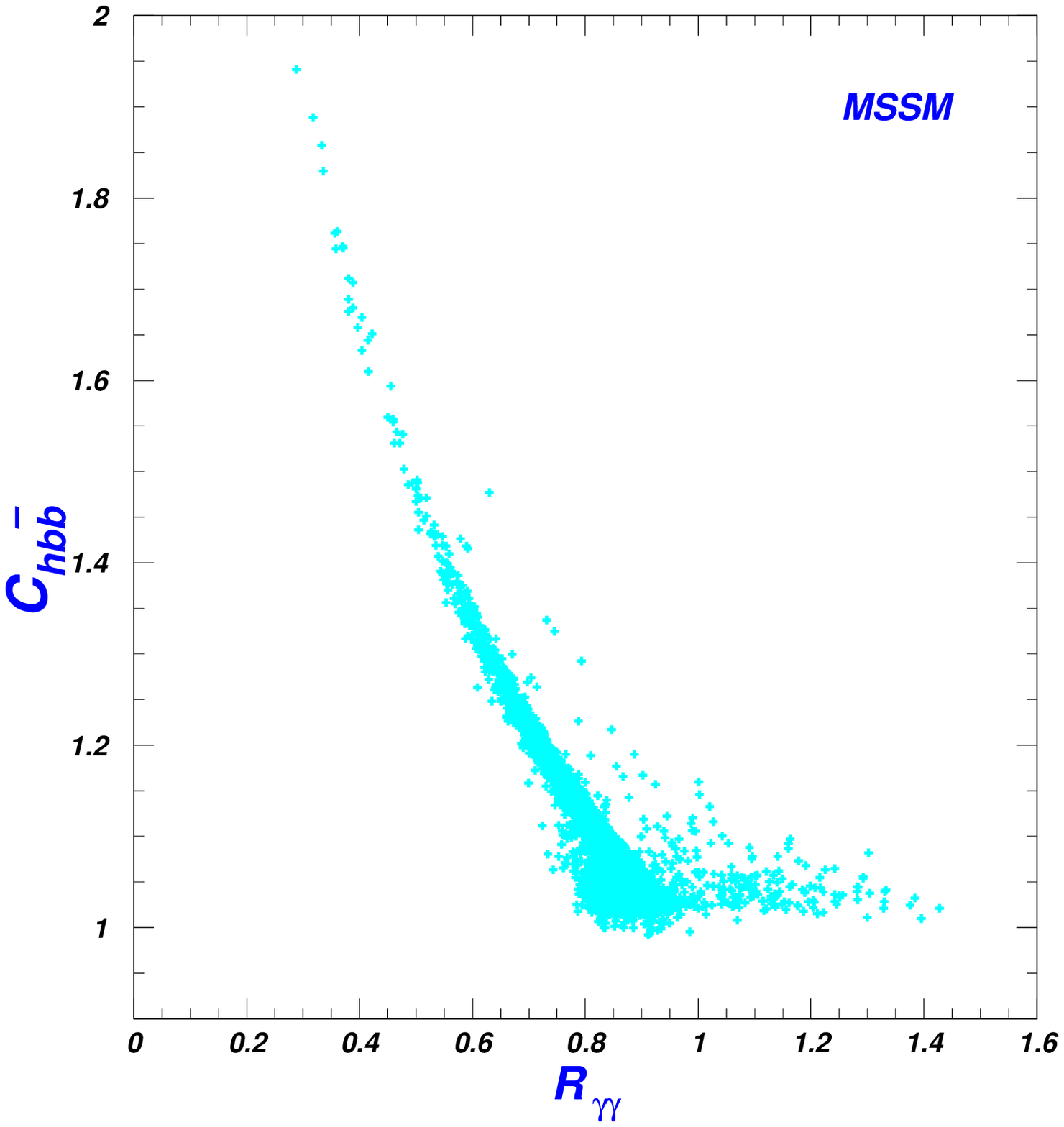}\hspace{0.2cm}
\includegraphics[width=7cm]{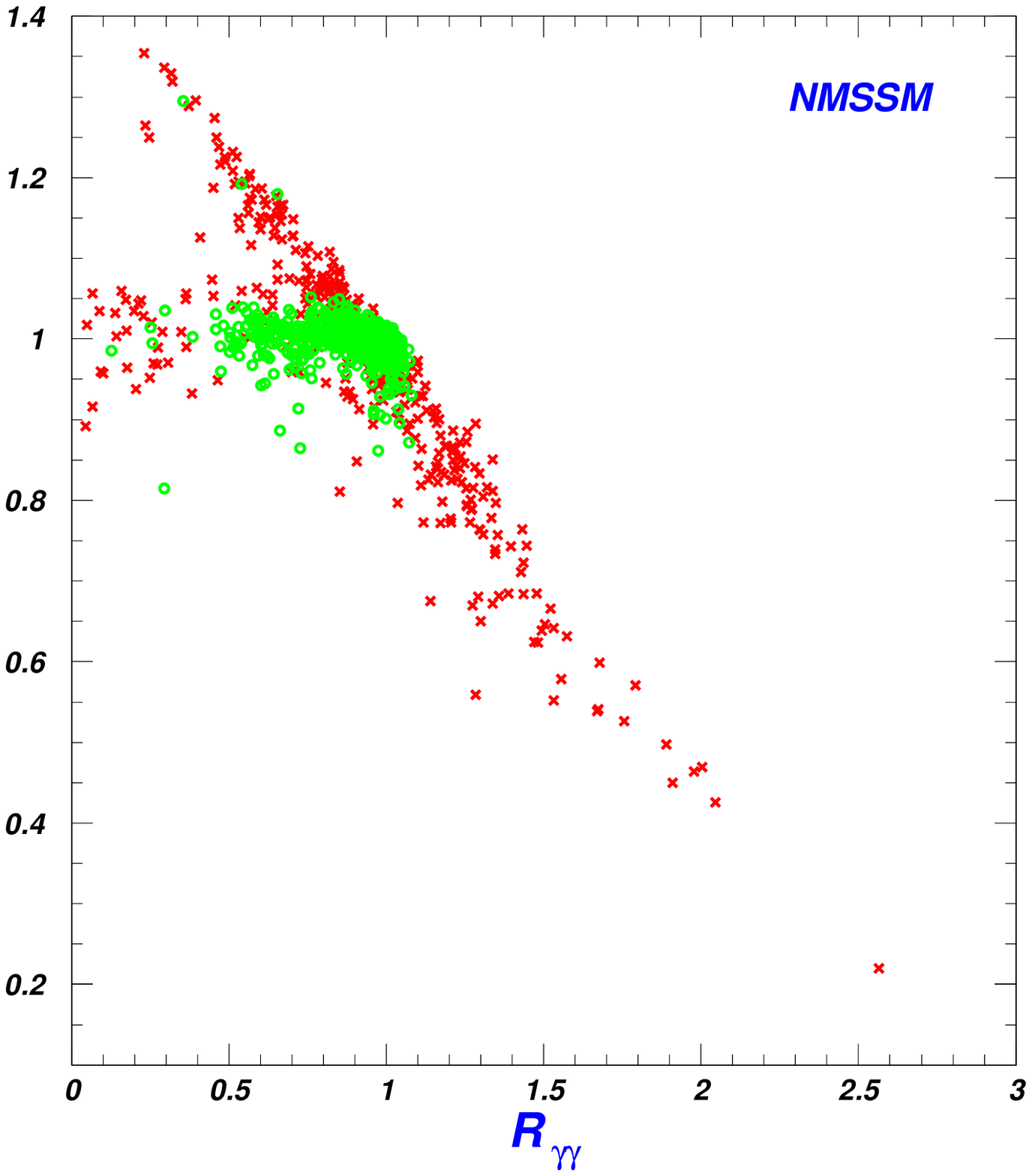}
\vspace*{-0.5cm}
\caption{Same as Fig.\ref{fig1}, but showing the dependence of the
di-photon signal rate $R_{\gamma\gamma}$
on the effective $h b\bar{b}$ coupling
$C_{h b\bar{b}}\equiv C^{\rm SUSY}_{h b\bar{b}}/C^{\rm SM}_{h b\bar{b}}$.
(taken for Ref.\cite{cao125})}
\label{fig5}
\end{figure}

Due to the clean background, the di-photon signal is crucial for searching for
the Higgs boson near 125 GeV. As discussed in the Sec.II, the signal rate is
relevant with the coupling $C_{h\gamma\gamma}$ and $C_{hgg}$ and the total width
of the SM-like Higgs boson. Both the coupling $C_{h\gamma\gamma}$ and $C_{hgg}$ are
affected by the contributions from the squark loops, especially the light top-squark loop,
so in the Fig.\ref{fig3} we give the relationship between the lighter top-squark mass
and the coupling $C_{h\gamma\gamma}$ and $C_{hgg}$, respectively. The figure shows that
the light $m_{\tilde t_1}$ may suppress the coupling $C_{hgg}$ significantly, especially
in the NMSSM. While the light top-squark has little effect on the coupling $C_{h\gamma\gamma}$
because there are additional contributions, as the Eq.(\ref{hgg})and Eq.(\ref{hgaga}) shown.
As analyzed in the Sec.II, light stau may enhance the coupling $C_{h\gamma\gamma}$, so in
Fig.\ref{fig4} we give the correlation between $m_{\tilde \tau_1}$ and the coupling
$C_{h\gamma\gamma}$ in the MSSM. The figure clearly shows that the coupling
$C_{h\gamma\gamma}$ can enhance to 1.25 for $m_{\tilde \tau_1}\sim$ 100 GeV.
Fig.4 also manifests that the enhancement of the coupling
$C_{h\gamma\gamma}$ corresponds to large $\mu\tan\beta$, which leads to large mixing.
These results exactly verifies the discussions in the Sec.II.

Since $h\to b\bar b$ is the main decay mode of the light Higgs boson, the total
width of the SM-like Higgs boson may be affected by the
effective $hb\bar b$ coupling $C_{hb\bar{b}}$, as discussed in \cite{di-photon}.
Under the effect of the combination $C_{hgg} C_{h\gamma \gamma}/C_{hb\bar{b}}$, the di-photon Higgs
signal rate may be either enhanced or suppressed, as shown in Fig.\ref{fig5}, which
also manifest that for the signal rate larger than 1, the effective $hb\bar b$ coupling
is enhanced a little in the MSSM, while it is suppressed significantly in the NMSSM.
Therefore, we can conclude that the reason for the enhancement in the signal rate is very different
between the MSSM and NMSSM. In the MSSM the enhancement of the signal is mainly due to the
enhancement of the coupling $C_{h\gamma\gamma}$, while in the NMSSM it is mainly due to the suppression
of the $hb\bar b$ coupling.

\begin{figure}[htb]
\centering
\includegraphics[width=7cm]{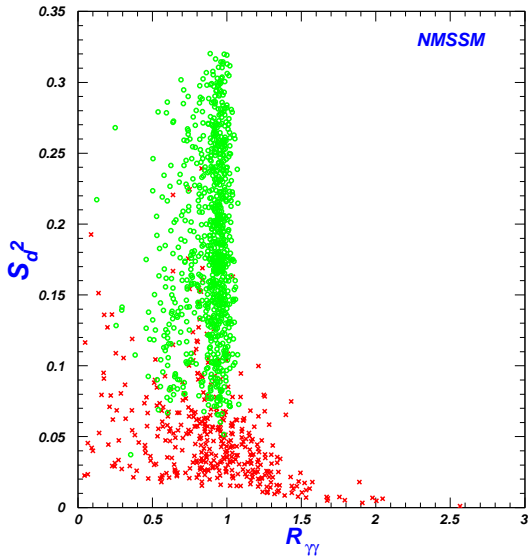}
\vspace*{-0.5cm}
\caption{Same as Fig.\ref{fig2}, showing the signal rate $R_{\gamma\gamma}$ versus
 $S_d^2$ with $S_d=C_{hb\bar b}\cos\beta$.}
\label{fig6}
\end{figure}

Due to the presence of the singlet field in the NMSSM, the doublet component in the
SM-like Higgs boson $h$ may be different from the case in the MSSM, which will affect
the coupling $hb\bar b$, and accordingly affect the total width of $h$. At the tree-level,
$C_{h b\bar{b}}=S_d/\cos\beta$. In Fig.\ref{fig6} we show the dependence of the
signal rate $R_{\gamma\gamma}$ on $S_d^2$. Obviously, for the signal rate larger than 1,
$S_d^2$ is usually very small, which leads to large suppression on the reduced coupling
$hb\bar b$. The figure also shows that the push-up case is more effective to enhance
the signal rate than the pull-down case. This is because the push-up case is easier
to realize the large mixing between the singlet field and the doublet field\cite{cao125}.

As the case in the NMSSM, nMSSM can also accommodate a 125 GeV SM-like Higgs\cite{di-photon}. However,
due to the peculiar property of the lightest neutralino $\tilde\chi_1^0$ in the nMSSM\cite{rarez},
the decay mode of the SM-like Higgs is very different from the case in the MSSM and NMSSM.
As discussed in the Sec.II, $h\to \tilde\chi_1^0\tilde\chi_1^0$ may be dominant 
over $h\to b\bar b$ in a major part of parameter space in the nMSSM \cite{di-photon,zhu}, which
induce a severe suppression on the di-photon Higgs signal.
Although the Higgs mass can be easily reach to 125 GeV, the di-photon signal
is not consistent with the LHC experiment. Therefore, the nMSSM may be excluded
by LHC experiment.
\begin{figure}[htbp]
\centering
\includegraphics[width=12cm]{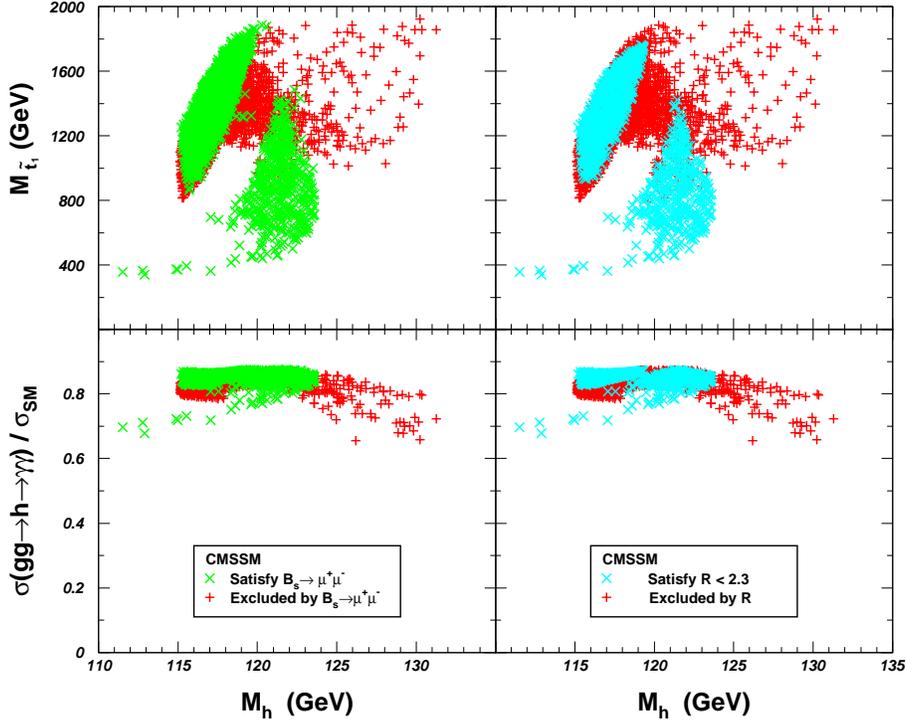}
\vspace*{-0.5cm}
\caption{The scatter plots of the surviving sample in the CMSSM, displayed in the planes
of the top-squark mass and the LHC di-photon rate versus the Higgs boson mass.
In the left frame, the crosses (red) denote the samples satisfying all
the constraints except $B_s\to\mu^+\mu^-$, and the times (green)
denotes those further satisfying the $Br(B_s\to\mu^+\mu^-)$
constraint. In the right frame, the crosses (red) are same as those
in the left frame, while the times (sky-blue) denote the samples
further satisfying the $R$ constraint.(taken for Ref.\cite{Cao:2011sn})}
\label{fig7}
\end{figure}
\begin{figure}[htbp]
\centering
\includegraphics[width=12cm]{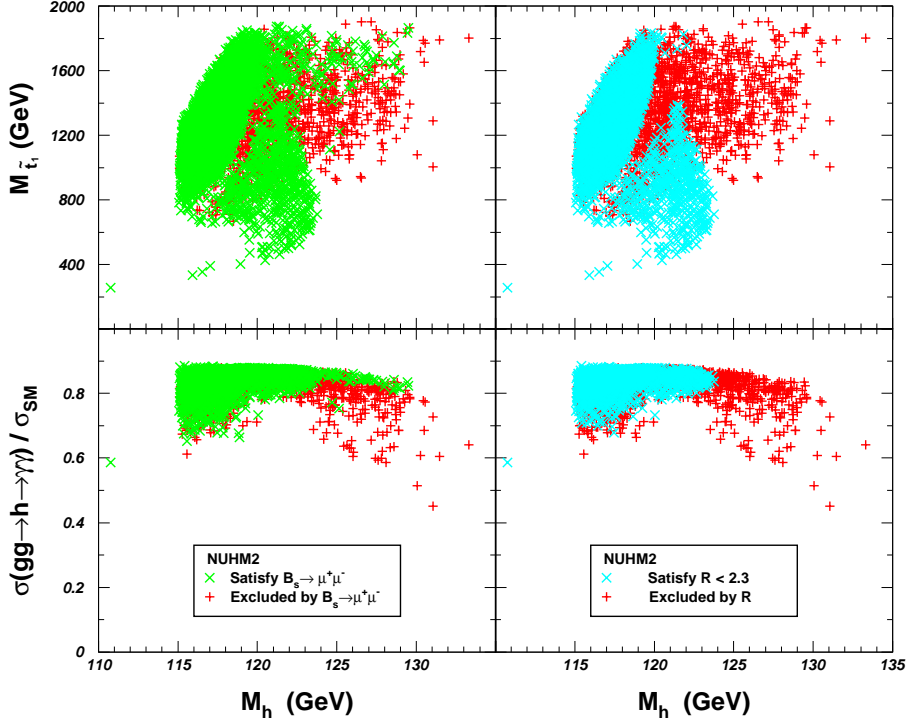}
\vspace*{-0.5cm}
\caption{Same as Fig.7, but for the NUHM2.(taken for Ref.\cite{Cao:2011sn})}
\label{fig8}
\end{figure}

We also considered the SM-like Higgs boson mass and its di-photon signal in the constrained MSSM
(including CMSSM and NUHM2) under various experimental constraints, 
especially the limits from $B_s\to \mu^+\mu^-$.
Because $Br(B_s\to \mu^+\mu^-)\propto\tan^6\beta/M_A^4$, so it may provide a rather strong
constraint on SUSY with large $\tan\beta$. Considering the large theoretical
uncertainties for the calculation of $Br(B_s\to \mu^+\mu^-)$, we use not only the LHCb data, but
also the double ratio of the purely leptonic decays defined as $R\equiv\frac{\eta}{\eta_{SM}}$
with $\eta\equiv\frac{Br(B_s\to\mu^+\mu^-)/Br(B_u\to\tau\nu_\tau)}
{Br(D_s\to\tau\nu_\tau)/Br(D\to\mu\nu_\mu)}$.
The surviving parameter space is plotted in Fig.\ref{fig7} for the CMSSM and Fig.\ref{fig8}
for the NUHM2. It shows that both the CMSSM and NUHM2 are hard to
realize a 125 GeV SM-like Higgs boson, and also the di-photon Higgs signal is suppressed
relative to the SM prediction due to the enhanced $h\bar bb$ coupling.
Therefore, the constrained MSSM may also be excluded by the LHC experiment.

\section{Conclusion}
In this work we briefly review our recent studies on a 125 GeV Higgs and its di-photon signal rate
in the MSSM, NMSSM, nMSSM and the constrained MSSM.
Under the current experimental constraints, we find: (i) the SM-like Higgs can easily
reach to 125 GeV in the MSSM, NMSSM and nMSSM, while it is hard to satisfy
in the constrained MSSM; (ii) the NMSSM may predict a lighter top-squark than the MSSM,
even as light as 100 GeV, which can ameliorate the fine-tuning problem;
(iii) the di-photon Higgs signal is suppressed
in the nMSSM and the constrained MSSM, but in a tiny corner of the parameter space in the MSSM and NMSSM,
it can be enhanced.

\section*{Acknowledgement}
The work is supported by the Startup Foundation for Doctors of Henan Normal University
under contract No.11108.

\end{document}